# Elfun18 – A collection of Matlab functions for the computation of Elliptical Integrals and Jacobian elliptic functions of real arguments.

Milan Batista

University of Ljubljana, Faculty of Maritime Studies and Transport, Portorož, Slovenia

milan.batista@fpp.uni-lj.si

**Abstract**

In the article, we outline the set of Matlab functions that enable the computation of elliptic Integrals and Jacobian elliptic functions for real arguments. Correctness, robustness, efficiency and accuracy of the functions are discussed in some details. An example from the elasticity theory illustrates use of the collection.

## 1 Introduction

Elfun18 is a collection of the Matlab functions that enable the computation of elliptic Integrals and Jacobian elliptic functions for real arguments. Its purpose is to provide a sufficiently comprehensive set of elliptic integrals and elliptic functions which one can meet in practice. Altogether the collection contains 75 different functions (see Appendix). In particular, the collection covers all elliptic integrals and Jacobian elliptic functions given in [1-5]. Most of these functions are for example included in computer algebra programs Maple and Mathematica and computational environment MATLAB. Also, several software libraries include elliptic integrals and elliptic functions (AMath/DAMath[1], BOOST[2], CERN[3], elliptic[4], DATAPLOT[5], SLATEC[6], Mathematics Source Library[7], MPMATH[8], MATHCW[9]). However, each of these programs and libraries has its limitations; either in a set of available functions or in the choice of function arguments. The reason for that is probably a maze of possibilities by which one can define elliptic integrals and functions. Thus the elliptic integrals can be given in three different forms [6]. Also, an argument of every elliptic integral or function is either the modular angle, the modulus or the parameter [3, 4]. To that, we add that for the Legendre form of elliptic integrals some these programs and libraries ignore the fact that these integrals are quasi-periodic functions and therefore give incorrect results.

[1] http://www.wolfgang-ehrhardt.de/amath_functions.html
[2] http://www.boost.org/doc/libs/1_53_0/libs/math/doc/sf_and_dist/html/math_toolkit/special/
[3] https://root.cern.ch/doc/v608/group__SpecFunc.html
[4] https://www.rdocumentation.org/packages/elliptic/versions/1.3-7
[5] http://www.itl.nist.gov/div898/software/dataplot/elli_fun.htm
[6] http://sdphca.ucsd.edu/slatec_source/TOC.htm
[7] http://www.mymathlib.com/functions/elliptic_integrals.html
http://www.mymathlib.com/functions/elliptic_functions.html
[8] http://docs.sympy.org/0.7.1/modules/mpmath/functions/elliptic.html
[9] http://ftp.math.utah.edu/pub/mathcw/

In the next section, we will outline the elliptic integrals and functions. In section 3 we will describe the collection structure and its functions. Section 4 we will discuss the collection correctness, robustness, efficiency and accuracy. Section 5 provides two examples from elasticity theory.

## 2 On elliptic integrals and elliptic functions

*General considerations.* As already mentioned an argument of any elliptic integral or any elliptic function is either the modular angle α, the modulus $k=\sin\alpha$, or the parameter $m=k^2$. Therefore three different notations are used [3]

$$Q(x\backslash\alpha),\quad Q(x,k),\quad Q(x|m)$$

where *Q* is an elliptic integral or elliptic function. We shall call these a-functions, k-functions, and m-functions.

Because our goal is a computation, we will as an elliptic argument use the parameter *m* [3]. The reasons are the following. First, using the modular angle reduce values of *k* to $-1\le k\le 1$ and values of *m* to $0\le m\le 1$. To obtain the values of an elliptic function for *k* and *m* outside these intervals, we must use reciprocal-modulus transformation formulas [1, 3]. Second, any elliptic integral and any Jacobian elliptic function is symmetric with respect to a real modulus *k*, that is,

$$Q(x,-k)=Q(x,k) \tag{1}$$

However, if the modulus is a pure imaginary number the symmetry does not hold. In this case, $(ik)^2=-k^2=m$, i.e. the parameter *m* is real. Therefore we have

$$Q(x,ik)=Q\left(x|-k^2\right)=Q(x|m) \tag{2}$$

Using the parameter *m* thus covers all real moduli and all pure imaginary moduli. Further, if we have a function $Q(x|m)$ then we can define

$$Q(x,k)\equiv Q\left(x|k^2\right) \tag{3}$$

$$Q(x\backslash\alpha)\equiv Q\left(x|\sin^2\alpha\right) \tag{4}$$

Thus $Q(x|m)$ covers $Q(x\backslash\alpha)$ and $Q(x,k)$ as a special case. All the functions in Elfun18 with *k* as an argument are defined in this way. However, we note that from a computational point of view $Q(x|m)$ reduce the domain of *k*. For example, in the double precision numerical model domain $10^{-308}<|k|<10^{308}$ is reduced to $10^{-154}<|k|<10^{154}$.

A similar situation is with the theta functions. Their argument is the nome *q*, $|q|<1$, which is for the real case given either by $q=\cos(\pi\tau)$, where $\tau$ is the half-period ration, or by $q=\exp\left(-\pi K'(m)/K(m)\right)$. So we have four possibilities

$$\vartheta(x,q),\quad \vartheta(x;\tau),\quad \vartheta(x,k),\quad \vartheta(x|m)$$

Currently, the collection covers only arguments *q* and *m*.

*Elliptic Integrals.* Historically, there is three kinds of elliptic integrals [1, 3, 7-10]:

- elliptic integrals of the first kind *F*,
- elliptic integrals of the second kind *E*, and
- elliptic integrals of the third kind $\Pi$.

Each of these classical integrals has associated so-called complete integral and complementary complete integral. Besides these, the integrals with particular names are integrals *B*, *C* and *D*. [2, 11] (Table 1).

The elliptic integrals depend on two arguments, except $\Pi$ which depends on three: one space argument and one (or two) elliptic arguments. With respect to the space argument any of the elliptic integrals can be written in three different normal forms [6]:

- Jacobi normal form with the argument *x*: $\mathbf{F}(x\,|\,m)$, $\mathbf{E}(x\,|\,m)$ and $\mathbf{\Pi}(x,\nu\,|\,m)$
- Legendre normal form with the argument $\phi$ (the amplitude): $F(\phi|m)$, $E(\phi|m)$ and $\Pi(\phi,\nu\,|\,m)$
- Jacobi second normal form with the argument *u*: $\mathcal{E}(u\,|\,m)$ and $\Lambda(u,\nu\,|\,m)$.

**Note 1**. In literature a usual notation of the Jacob's second form of the elliptic integral of the second kind is $E(\mathrm{am}\,x\,|\,m)$. Notation $\mathcal{E}(u\,|\,m)$ is introduced in [5]

**Note 2.** The original Jacobi's form of elliptic integral of the third kind is $\int_0^u \frac{\kappa^2 \mathrm{sn}a\,\mathrm{cn}a\,\mathrm{dn}a\,\mathrm{sn}^2 t}{1-\kappa^2 \mathrm{sn}^2 a\,\mathrm{sn}^2 t}dt$ [7, 9, 12, 13]. Notation $\Lambda(u,\nu\,|\,m)$ was introduced by Lawden [14]. This notation is not related to Heuman's $\Lambda(\phi,\beta\,|\,m)$ [15].

From Jacobi's form of the integrals, we obtain Legendre form of elliptic integrals by substitution $x=\sin\phi$, and the Jacobi's second form by substitution $x=\mathrm{sn}(u\,|\,m)$. From Legendre form of elliptic integral, we obtain Jacobi's second form by substitution $\phi=\mathrm{am}(u\,|\,m)$. Here sn is Jacobi function and am is Jacobi amplitude function. We note that unlike integrals are given by Jacobi's first form, the integrals in Legendre form and Jacobi's second form are quasiperiodic functions. Besides the elliptic integral of the third kind, there is also the Heuman's form [15]. However only the complete form of this integral denoted as $\Lambda_0(\beta\,|\,m)$ is used.

The collection also contains the functions $Z(u\,|\,m)$ and $\Omega(u,\nu\,|\,m)$. The first is the periodic part of function $\mathcal{E}(u\,|\,m)$; the second is the periodic part of function $\Lambda(u,\nu\,|\,m)$ [16]. We note that Legendre form of these functions $Z(\phi|m)$ and $\Omega(\phi,\nu\,|\,m)$ (which are also included in the collection) are not the periodic part of functions $E(\phi|m)$ and $\Pi(\phi,\nu\,|\,m)$.

*Jacobian elliptic functions.* Jacobian elliptic functions are inverse of the elliptic integrals [11]. The basic connection between elliptic integrals and Jacobin elliptic functions are the following

$$x=F(\phi\,|\,m),\quad \phi=\mathrm{am}(x\,|\,m) \tag{5}$$

That is, Jacobi's amplitude function *am* is inverse of Legendre form of the elliptic integral of the first kind *F*. The three basic Jacobian elliptic functions are then

$$\mathrm{sn}(x\,|\,m)=\sin\phi\,,\quad \mathrm{cn}(x\,|\,m)=\cos\phi\,,\quad \mathrm{dn}(x\,|\,m)=\sqrt{1-m\sin^2\phi} \tag{6}$$

Besides these three elliptic functions, there are also nine Glaisher elliptic functions which are defined as quotients of these functions [4, 5]. In the collection, all Jacobian elliptic functions are calculated by procedure *sncndn* [17, 18]. The procedure is based on Gauss transformation and solves $x=\mathbf{F}\left(\mathrm{sn}\left(x\,|\,m\right),m\right)$ for $\mathrm{sn}\left(x\,|\,m\right)$.

The inverse of Jacobian elliptic functions are but a special case of the elliptic integral of the first kind [3].

## 3 Collection description

The functions in the collection are available in two levels:

- low-level functions with scalar arguments
- higher level functions with matrix arguments

It is assumed that the input arguments of low-level functions are real scalars without check. Actual computation is conducted by the low-level functions which as input has the parameter *m* (m-functions). Most of these computational functions are based on Matlab translations of Algol procedures from [17, 19, 20], Fortran functions from [18, 21], and Pascal procedures from [22]. More precisely, for computation of elliptic integrals and inverse of Jacobian elliptic functions either Bulirsch's integrals *el1*, *el2*, *el3* or Carlson's integrals *rc*, *rd*, *rg*, *rj*, are used [1]. The core function for computation of Jacobian elliptic function is procedure *sncndn* from [1]. Most of the low-level functions which use the module *k* as input argument are wrappers, i.e. the functions which calls appropriate low-level m-function by setting $m=k^2$.

All high-level functions are wrappers which mimics an elemental behavior of a function by calling either of the functions *ufun1*, *ufun2*, *ufun3*, *ufun4*. Here the term *elemental* is browed from Fortran, i.e. it means that high level functions may be called with matrix arguments of the same size (or any of them can be scalar) in which case an coresponded low-level function is applied element-wise, with a conforming matrix return value. All higher level functions check its input data; matrix class and a number of arguments.

*Naming convention*: All high-level functions use Maple style naming, i.e. the functions begin with an uppercase letter. In this way, they cannot be a mismatch with Matlab symbolic toolbox functions which begin with a lowercase letter. Each function is available either with the modulus *k* or parameter *m* as an argument. In the later case, the function name start with letter 'm'. Incomplete elliptic integrals are given either in Jacobi form, Legendre form or Jacobi's second form. Legendre form begins with letter 'p' (from the argument which is 'phi'). For the Jacobi's second form of elliptic integrals we use special names: JacobiEpsilon and JacobiLambda.

We enhance the above description with examples. If the modulus *k* is used as an argument then the low-level name for Jacobian elliptic function *sn* is 'jsn' and its higher level name is 'JacobiSN'. If the parameter *m* is used as argument then the names are 'mjsn' and 'mJacobiSN'. Low-level name for a function which calculates the Jacobi form of elliptic integral of first kind with *k* as argument is 'elF' and the name of the higher level function is EllipticF. If the parameter *m* is used as an argument then the names are 'melF' and mEllipticF. If Legendre form of integral is used then the names are either 'pelF' and pEllipticF or 'mpelF' and mpEllipticF.

The names of some function are not standard. Thus for convenience, we named $\Lambda(u,\nu|m)$ and $\Omega(\phi,\nu|m)$ as JacobiLambda and JacobiOmega, trough Jacobi does not use these functions [13].

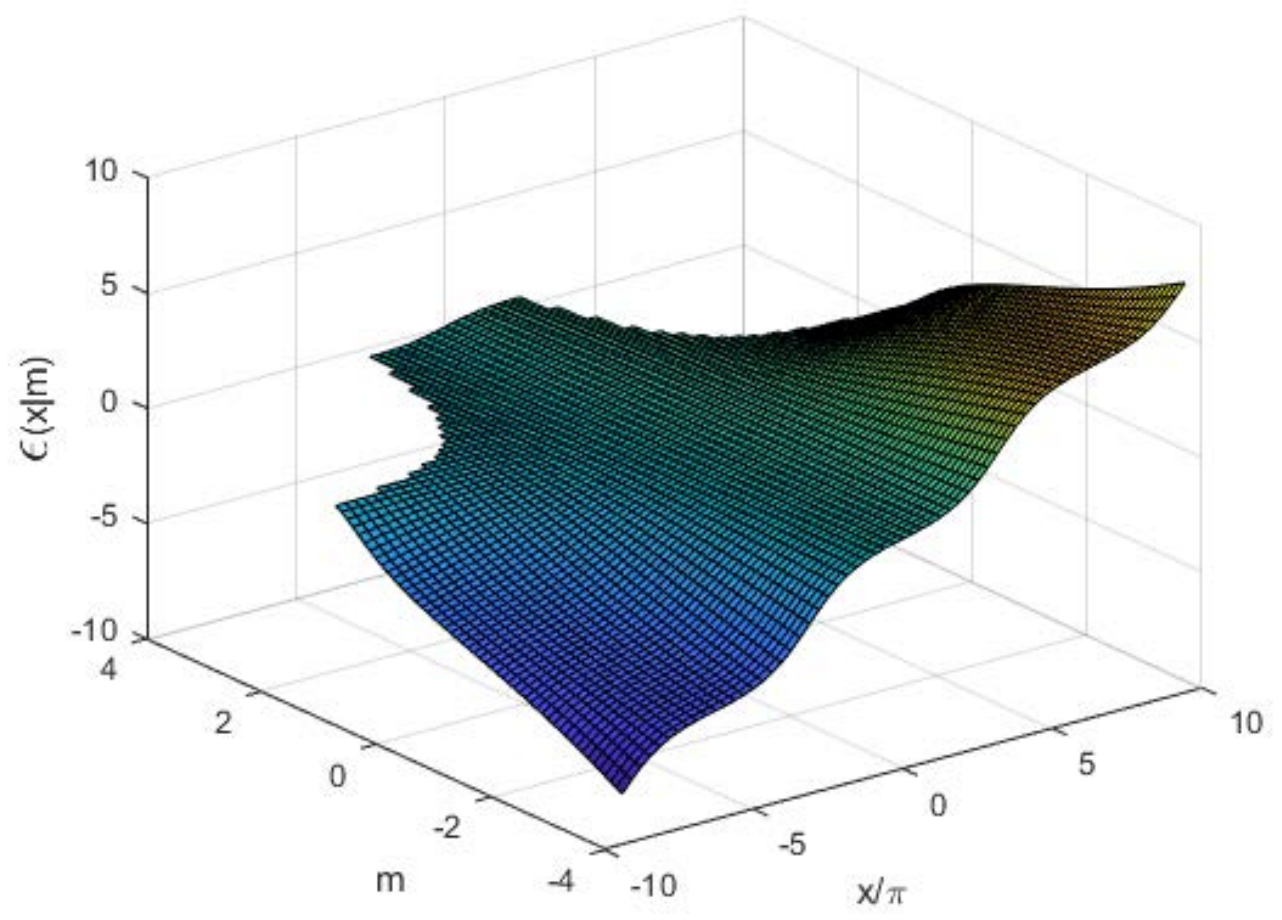

**Figure 1.** Jacob's epsilon function

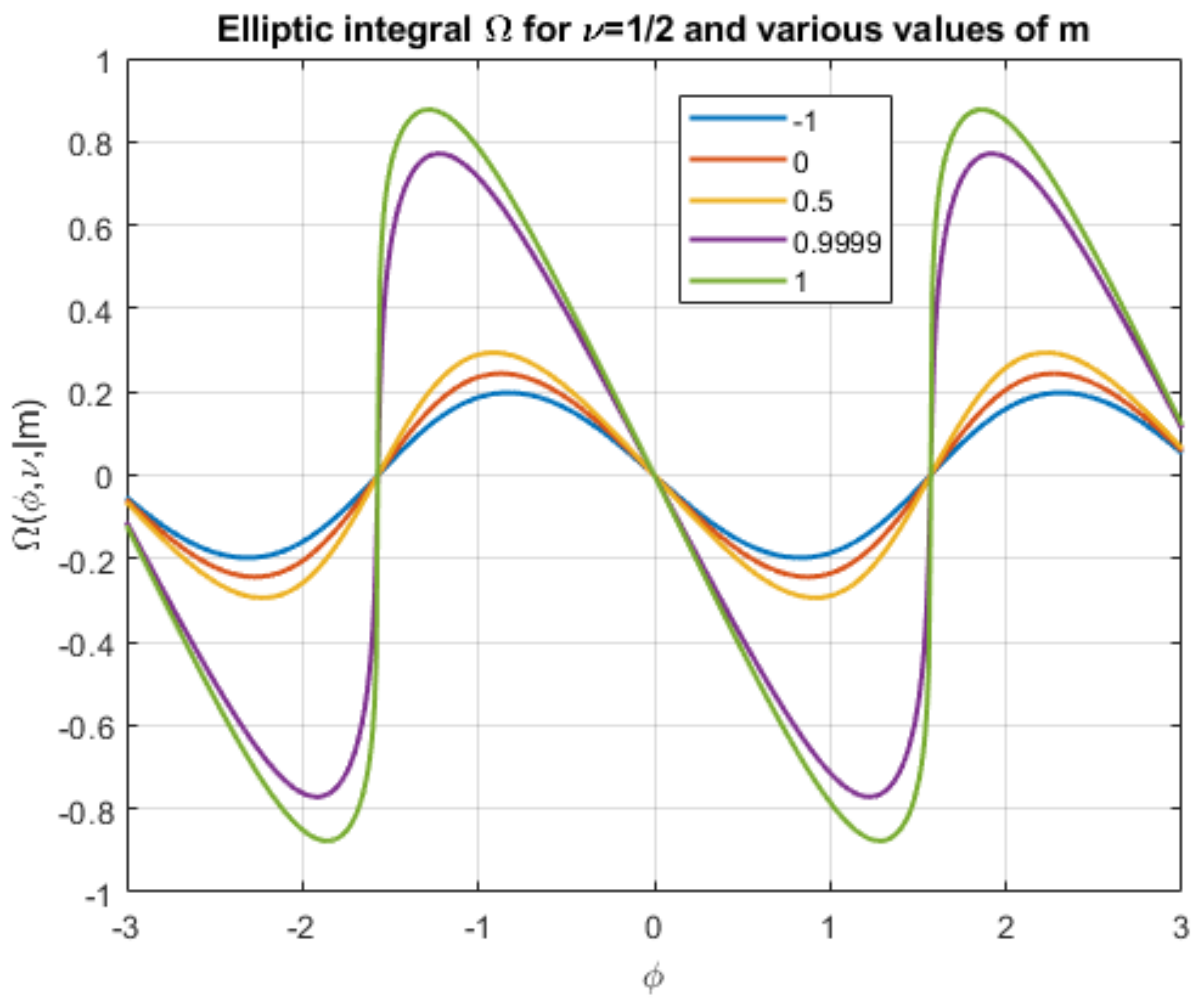

**Figure 2.** Legendre form of the omega function

*4.1 Correctness*

From the functions in the collection, we expect that for real arguments they return either a real number (including Inf) or NaN. To assure that any computational function checks if its actual arguments belong to the domain where the function values are real numbers. If an argument falls out of the domain the function return NaN without warning. It is thus user responsibility to check obtained results. For example, the Jacobi form of elliptic integrals has real values when $1-x^2\geq 0$ and $1-mx^2\geq 0$. From this, it follows that when $m\leq 1$ then $|x|\leq 1$, and if $m>1$ then $|x|<1/\sqrt{m}$. The parameter $m$ is thus any real number while the domain of $x$ depends on $m$. For the second example, we consider complete elliptic integral $K(m)$. In the function melK its domain is set to $-\infty<m\leq 1$ because for $m>1$ the values of the integral are complex.

Some care was also taken to design the functions which intercept their poles and their limits for infinite arguments. Thus for example $K(1)=\infty$ and $K(-\infty)=0$. Therefore for these arguments, the collection function melK gives $\text{melK}(1)=\text{Inf}$ and $\text{melK}(-\text{Inf})=0$. This approach differs from typical implementations where NaN is returned if for example $m=-\infty$.

To assure that for $m=k^2$ the *m*-functions give the same value as the *k*-functions all *k*-functions are defined by (3), i.e., all *k*-functions are wrappers. Also, if a function $f(x)$ is symmetric or asymmetric then for $x<0$ it is evaluated as $f(-x)=f(x)$ or $f(-x)=-f(x)$.

### *4.2 Robustness*

In the present collection, the number and type of input data are checked in higher-level routines. Low-level routines intercept inputs which are NaN and out of function domain.

All computational routines were tested for very small and for huge arguments. Some additional checks are added to the function which use successive Landen's transformation to prevent them from hanging if any variable become NaN.

### *4.3 Efficiency*

As was already stated the functions in the collection mimics the elemental-behavior on the matrix arguments i.e. the code is not vectorized. This has an impact to functions efficiency if a vector or matrix arguments are used. From Table 1-2 we can see that vectorised MATLAB function are on matrix argument two to five times faster than present functions. However, this is not critical because as we can see from Table 3 the computation time for a million of function evolutions is a matter of half of the second. Also from the Table1-3, we can also see that the present functions are faster from Matlab functions in the case when a function is called with a scalar argument. As expected the present functions are much faster than elliptic functions from Matlab symbolic toolbox.

**Table 1.** Execution time for computation of the complete elliptic integral of the first kind.

| | $10^6$ individual call | | Vector argument $10^4$ | | Matrix argument 1000 x1000 | |
|---|---|---|---|---|---|---|
| | Time s | Ratio | Time s | Ratio | Time s | Ratio |
| melK/mEllipticK | 0.524 | 1 | 1.270 | 1 | 1.273 | 1 |
| elipke | 3.235 | 6.17 | 0.230 | 0.18 | 0.210 | 0.17 |

**Table2.** Execution time for computation of Jacobian elliptic function *sn*

| | $10^3$ individual calls | | Vector argument 1000 | | Matrix argument 100 x100 | |
|---|---|---|---|---|---|---|
| | Time s | ratio | Time s | Ratio | Time s | Ratio |
| mjsn/mJacobiSN | 0.0025 | 1 | 0.0028 | 1 | 0.0226 | 1 |
| ellipj | 0.0286 | 11.3 | 0.0019 | 0.68 | 0.0102 | 0.45 |
| jacobiSN | 15.892 | 6269 | 0.896 | 324 | 7.981 | 352 |

**Table3.** Solution of $2E(k)=K(k)$ by Matlab function fzero

| | Average Time ms | Ratio |
|---|---|---|
| elK/elF | 4.3 | 1 |
| ellipke | 5.1 | 1.18 |

*4.5 Accuracy*

The accuracy of the functions in the collection was testate in two ways: with comparison with MATLAB elliptic functions from the symbolic toolbox and by testing various functional identities. Examples of the results of such tests are shown in Tables 4-5. These tests show that functions are accurate to about 9 digits for $|m|<10^3$ and $|x|<10^4$.

**Table 4**. Accuracy test with random input. Comparison with MATLAB symbolic toolbox functions ellipticE, ellipticF, ellipticPi. Number of runs 1000. MEA is maximal relative error, MRE is maximal relative error, RMS is root mean square error. $eps \approx 2.22\times10^{-16}$

| func. | min(\|x\|) | max(\|x\|) | min(m) | max(m) | MAE/eps | MRE/eps | RMS/eps |
|---|---|---|---|---|---|---|---|
| mpEllipticE | 4.3E-04 | 5.0E-01 | -5.0E-01 | 5.0E-01 | 1.5 | 4 | 1.2 |
| mpEllipticE | 1.0E-04 | 1.6E+01 | -1.6E+01 | 1.6E+01 | 158 | 51 | 2.1 |
| mpEllipticE | 1.5E-06 | 5.0E+02 | -4.2E+02 | 4.2E+02 | 5.45E+04 | 688 | 37.5 |
| mpEllipticE | 1.9E-07 | 1.6E+04 | -5.1E+03 | 5.3E+03 | 3.80E+06 | 973 | 61.3 |
| mpEllipticF | 8.2E-04 | 5.0E-01 | -5.0E-01 | 5.0E-01 | 1 | 3 | 0.9 |
| mpEllipticF | 2.4E-05 | 1.6E+01 | -1.6E+01 | 1.5E+01 | 48 | 3 | 1.0 |
| mpEllipticF | 5.9E-06 | 5.0E+02 | -1.7E+02 | 3.7E+02 | 768 | 3.5 | 0.9 |
| mpEllipticF | 3.4E-08 | 1.6E+04 | -1.0E+04 | 1.6E+04 | 4.10E+04 | 3 | 0.9 |
| mpEllipticPi | 5.1E-05 | 5.0E-01 | -5.0E-01 | 5.0E-01 | 5 | 3 | 1.1 |
| mpEllipticPi | 1.2E-05 | 1.6E+01 | -1.6E+01 | 9.9E-01 | 32 | 9 | 1.2 |
| mpEllipticPi | 4.9E-06 | 4.8E+02 | -5.0E+02 | 1.0E+00 | 32 | 25 | 2.5 |
| mpEllipticPi | 3.4E-08 | 1.3E+04 | -6.5E+03 | 9.9E-01 | 64 | 112 | 8.8 |

**Table 5**. Accuracy test. $x=0.23$ , $k=0.999$ . $N=10^5$

| Function | f(x) | f(x+N*K(k)) | Diff. |
|---|---|---|---|
| jcd | 0.999946 | 0.999946 | 1.42E-12 |
| jcn | 0.974120 | 0.974120 | 6.56E-10 |
| jcs | 4.309650 | 4.309650 | 5.68E-08 |
| jdc | 1.000050 | 1.000050 | -1.42E-12 |
| jdn | 0.974172 | 0.974172 | 6.54E-10 |
| jds | 4.309880 | 4.309880 | 5.68E-08 |
| jnc | 1.026570 | 1.026570 | -6.91E-10 |
| jnd | 1.026510 | 1.026510 | -6.89E-10 |
| jns | 4.424150 | 4.424150 | 5.53E-08 |
| jsc | 0.232037 | 0.232037 | -3.06E-09 |
| jsd | 0.232025 | 0.232025 | -3.06E-09 |
| jsn | 0.226032 | 0.226032 | -2.83E-09 |
| jzeta | 0.174671 | 0.174671 | 1.53E-11 |

## 5 Examples

As simple application of present formulas consider Euler's flexural elastica which has parametric form [23, 24]

$$x(u)=\frac{2}{\omega}\left[\mathcal{E}(\omega s+C,k)-\mathcal{E}(C,k)\right]-s,\; y(s)=\frac{2k}{\omega}\left[\mathrm{cn}(C,k)-\mathrm{cn}(\omega s+C,k)\right] \quad (7)$$

where $\omega$ is load parameter and $C$ is constant. The following program produces the shapes of the elstica shown on Fig 3

```
% Flexural elastica

% Data
omega = 5;
cc     = 1;
kk     = [0.1,0.2,0.3,0.4,0.5,0.6,0.7,0.8,0.9];

% Plot
figure
hold on
s = 0:0.01:1;
C = cc*ones(size(s));
for n = 1:length(kk)
    k = kk(n);
    x = -s + 2*(JacobiEpsilon(omega*s + C,k) - JacobiEpsilon(C,k))/omega;
    y = 2*k*(JacobiCN(C,k) - JacobiCN(omega*s + C,k))/omega;
    plot(x,y,'LineWidth',2)
end
legend('0.1','0.2','.3','0.4','0.5','0.6','0.7','0.8','0.9',...
    'Location','best')
axis equal
```

```
grid on
hold off
```

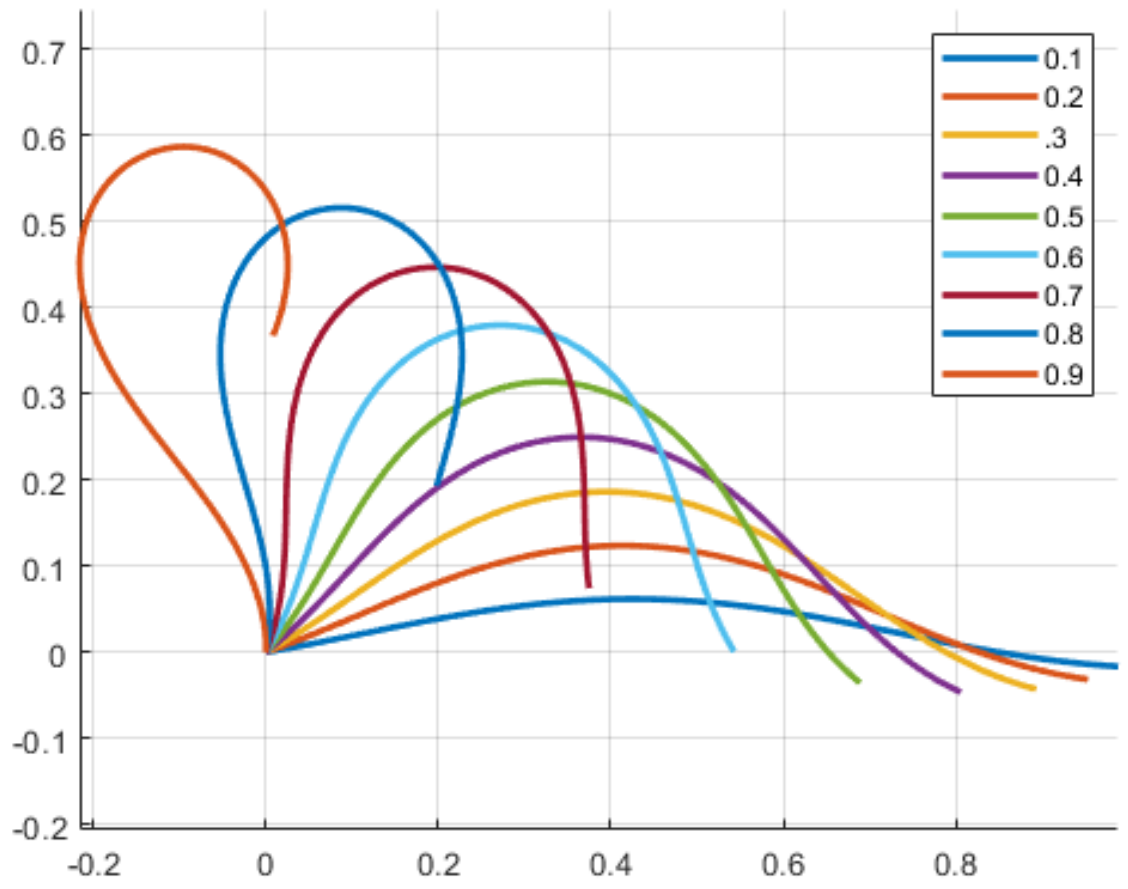

**Figure 3.** Flexural elastica for $\omega = 5$, $C = 1$ and different values of $k$

As a second example, we consider finite-strain elastic cantilever under follower force from [24]. The coordinates $X$, Y of deformed cantilever base curve are

$$X = x\cos\alpha + y\sin\alpha\,, \quad Y = -x\sin\alpha + y\cos\alpha \tag{8}$$

where

$$\begin{aligned} x = &-\left[\frac{(1-\nu)\omega^2}{2\lambda^2} + \frac{\tilde{\omega}^2}{\omega^2}\right]s + \frac{2\tilde{\omega}}{\omega^2}\Bigg\{\mathcal{E}(\tilde{\omega}s + C, k) - \mathcal{E}(C,k) \\ &-m^2\left[\frac{\mathrm{sn}(\tilde{\omega}s+C,\tilde{k})\mathrm{cn}(\tilde{\omega}s+C,\tilde{k})\mathrm{dn}(\tilde{\omega}s+C,\tilde{k})}{1+m^2\mathrm{cn}^2(\tilde{\omega}s+C,\tilde{k})} - \frac{\mathrm{sn}(C,\tilde{k})\mathrm{cn}(C,\tilde{k})\mathrm{dn}(C,\tilde{k})}{1+m^2\mathrm{cn}^2(C,\tilde{k})}\right]\Bigg\} \end{aligned} \tag{9}$$

$$y = \frac{2k\tilde{\omega}\sqrt{1+m^2}}{\omega^2}\left[\frac{\mathrm{cn}(C,\tilde{k})}{1+m^2\mathrm{cn}^2(C,\tilde{k})} - \frac{\mathrm{cn}(\tilde{\omega}s+C,\tilde{k})}{1+m^2\mathrm{cn}^2(\tilde{\omega}s+C,\tilde{k})}\right]$$

and

$$\tilde{\omega} \equiv \omega\sqrt{1 + \frac{\nu\omega^2}{\lambda^2}(2k^2-1)}\,, \quad m^2 \equiv \frac{\frac{\nu\omega^2}{\lambda^2}k^2}{1 - \frac{\nu\omega^2}{\lambda^2}(1-k^2)}\,, \quad \tilde{k}^2 \equiv \frac{k^2+m^2}{1+m^2} \tag{10}$$

The non-dimensional parameters in the above equations are the load parameter ω, the generalized slenderness ratio $\lambda$ and the stiffness ratio. For cantilever under follower force we have

$$k^2 = \sin^2\frac{\psi_1}{2} \tag{11}$$

$$\alpha = 2\sin^{-1}\left(k\frac{\operatorname{sn}(C,\tilde{k})}{\sqrt{1+m^2\operatorname{cn}^2(C,\tilde{k})}}\right) \tag{12}$$

$$C = -\tilde{\omega} + K(\tilde{k}) \tag{13}$$

where $\psi_1$ is given the clockwise angle between the direction of force and inward normal to the cantilever deformed cross section at the free end. When the cantilever is shearless i.e. when $\nu = 1$, its deformed length $L$ is given by

$$L = 1 - \left(1+\frac{2k^2}{m^2}\right)\frac{\omega^2}{\lambda^2} + \frac{2k^2}{m^2\tilde{\omega}}\frac{\omega^2}{\lambda^2}\left[\Lambda\left(\tilde{\omega}+C,\frac{m^2}{1+m^2},\tilde{k}\right) - \Lambda\left(C,\frac{m^2}{1+m^2},\tilde{k}\right)\right] \tag{14}$$

The program which implements these formulas is given in Appendix B. The shape of deformed cantilever shown in Fig 4 match with the shape in Fig 3c [24].

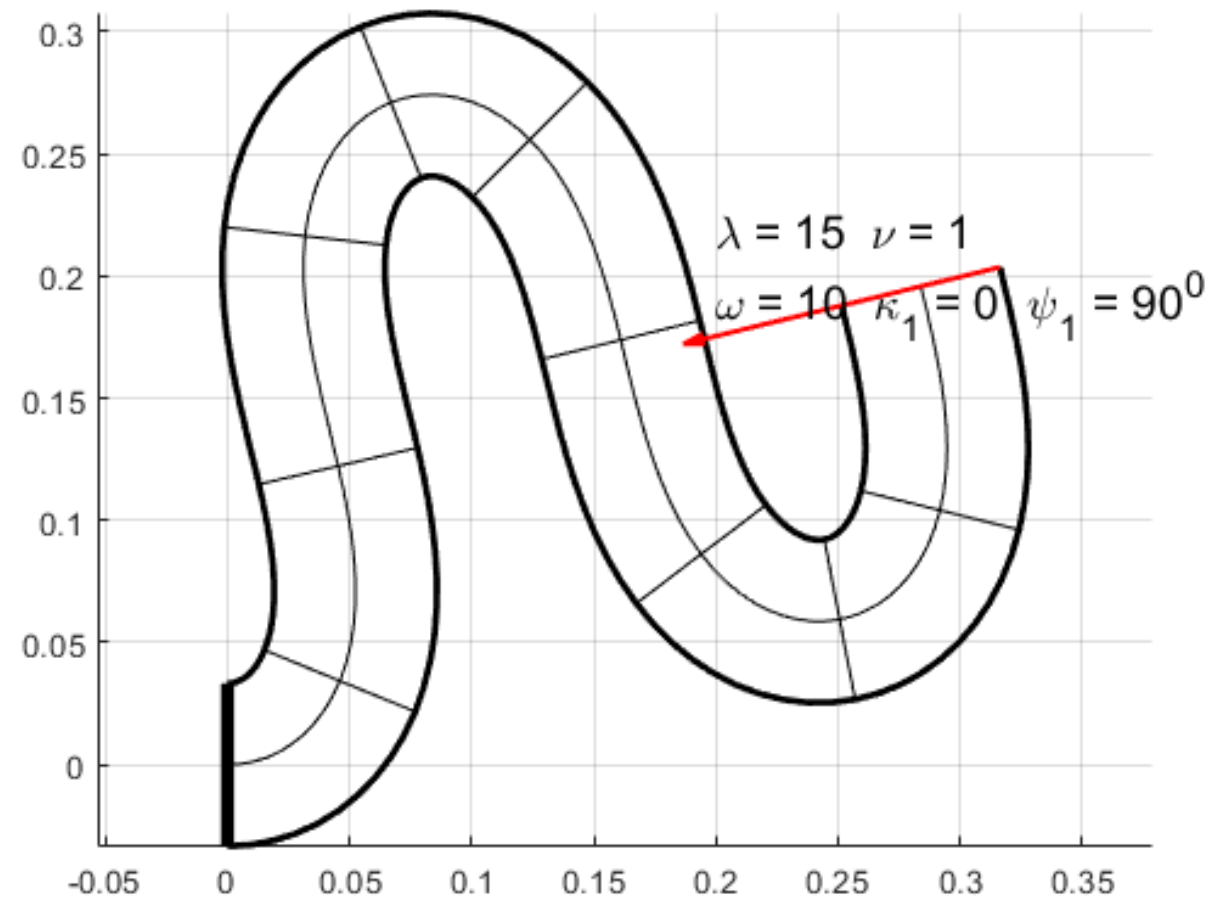

**Figure 4**. Shearless cantilever under follower force. The length of the deformed base curve 0.78555198 which agree with the value given in Table 1 [24].

## 7 Conclusion

The described collection contains a complete set of Matlab functions for the calculation of real elliptic integrals and real Jacobian elliptic functions in their all possible forms. The functions were tested for correctness, robustness, efficiency, and accuracy. The collection is freely available from https://www.mathworks.com/matlabcentral/fileexchange/65915-elfun18 . The reference manual with the examples is available from https://www.researchgate.net/publication/323399074_elfun18_-_Elliptic_Integrals_Jacobian_Elliptic_Functions_and_Theta_Functions_-_Reference_Manual_v11.

**Appendix A.** Contents of the collection

**Table.** The notation used for elliptic integrals an functions

| Notation | Description | Relations |
|---|---|---|
| $k$ | modulus | |
| $k'$ | complementary modulus | $k^2+k'^2=1$ |
| $m$ | parameter | $m=k^2$ |
| $m'$ | complementary parameter | $m+m'=1$ |
| $q$ | nome | $q=\exp(-\pi K'/K)$ |
| $q'$ | Complementary nome | $q_1=\exp(-\pi K/K')$ |
| $x$ | | $x=\sin\phi=\mathrm{sn}(u\mid m)$ |
| $\alpha$ | modular angle | $m=\sin^2\alpha$ , $k=\sin\alpha$ |
| $\beta$ | characteristic angle | $\nu=\sin\beta$ |
| $\nu$ | characteristic | |
| $\phi$ | amplitude | $\phi=\mathrm{am}(u\mid m)$ |

**Table A1.** Bulirsch's elliptic integrals

| Function | Matrix arguments | Scalar arguments |
|---|---|---|
| $el1(x,k')\equiv\int_0^x\frac{dt}{\sqrt{(1+t^2)(1+k'^2t^2)}}$ | BulirschEL1(X,KC) | el1(x,kc) |
| $el2(x,k',a,b)\equiv\int_0^x\frac{(a+bt^2)dt}{(1+t^2)\sqrt{(1+t^2)(1+k'^2t^2)}}$ | BulirschEL2(X,KC,A,B) | el2(x,kc,a,b) |
| $el3(x,k',p)\equiv\int_0^x\frac{(1+t^2)dt}{(1+pt^2)\sqrt{(1+t^2)(1+k'^2t^2)}}$ | BulirschEL3(X,KC,P) | el3(x,kc,p) |
| $cel1(k')\equiv el1(\infty,k')$ | BulirschCEL1(kc) | cel1(kc) |
| $cel2(k',a,b)\equiv el2(\infty,k',a,b)$ | BulirschCEL2(KC,A,B) | cel2(kc,a,b) |
| $cel3(k',p)\equiv el3(\infty,k',p)$ | BulirschCEL3(KC,P) | cel3(kc,p) |
| $cel(k',p,a,b)\equiv\int_0^\infty\frac{(a+bt^2)dt}{(1+pt^2)\sqrt{(1+t^2)(1+k'^2t^2)}}$ | BulirschCEL(KC,P,A,B) | cel(kc,p,a,b) |

**Table A2.** Carlson's elliptic integrals.

| Function | Matrix argument | Scalar arguments |
|---|---|---|
| $R_C(x,y) \equiv R_F(x,y,y)$ | CarlsonRC(X,Y) | rc(x,y) |
| $R_D(x,y,z) \equiv R_J(x,y,z,z)$ | CarlsonRD(X,Y,Z) | rd(x,y,z) |
| $R_F(x,y,z) \equiv \frac{1}{2}\int_0^\infty \frac{dt}{\sqrt{(t+x)(t+y)(t+z)}}$ | CarlsonRF(X,Y,Z) | rf(x,y,z) |
| $R_G(x,y,z) \equiv \frac{1}{4}\int_0^\infty \frac{1}{\sqrt{(t+x)(t+y)(t+z)}}\left(\frac{x}{t+x}+\frac{y}{t+y}+\frac{z}{t+z}\right)t\,dt$ | CarlsonRG(X,Y,Z) | rg(x,y,z) |
| $R_J(x,y,z,p) \equiv \frac{3}{2}\int_0^\infty \frac{dt}{(t+p)\sqrt{(t+x)(t+y)(t+z)}}$ | CarlsonRJ(X,Y,Z,P) | rj(x,y,z,p) |

**Table A3.** Incomplete elliptic integrals $-\infty < m < \infty$

| Jacobi form | Matrix argument | Scalar argument |
|---|---|---|
| $\mathbf{B}(x\mid m)\equiv\int_0^x \frac{(1-t^2)\,dt}{\sqrt{(1-t^2)(1-mt^2)}}$ | mEllipticB(X,M) | melB(x,m) |
| $\mathbf{D}(x\mid m)\equiv\int_0^x \frac{t^2\,dt}{\sqrt{(1-t^2)(1-mt^2)}}$ | mEllipticD(X,M) | meld(x,m) |
| $\mathbf{E}(x\mid m)\equiv\int_0^x \sqrt{\frac{1-mt^2}{1-t^2}}\,dt$ | mEllipticE(X,M) | melE(x,m) |
| $\mathbf{F}(x\mid m)\equiv\int_0^x \frac{dt}{\sqrt{(1-t^2)(1-mt^2)}}$ | mEllipticF(X,M) | melF(x,m) |
| $\mathbf{\Pi}(x,\nu\mid m)\equiv\int_0^x \frac{dt}{(1-\nu t^2)\sqrt{(1-t^2)(1-mt^2)}}$ | mEllipticPi(X,NU,M) | melPi(x,m) |
| Legendre form | | |
| $B(\phi\mid m)\int_0^\phi \frac{\cos^2\theta\,d\theta}{\sqrt{1-m\sin^2\theta}}$ | mpEllipticB(PHI,M) | mpelB(phi,m) |
| $D(\phi\mid m)\equiv\int_0^\phi \frac{\sin^2\theta\,d\theta}{\sqrt{1-m\sin^2\theta}}$ | mpEllipticD(PHI,M) | mpelD(phi,m) |
| $E(\phi\mid m)\equiv\int_0^\phi \sqrt{1-m\sin^2\theta}\,d\theta$ | mpEllipticE(PHI,M) | mpelE(phi,m) |
| $F(\phi\mid m)\equiv\int_0^\phi \frac{d\theta}{\sqrt{1-m\sin^2\theta}}$ | mpEllipticF(PHI,M) | mpelF(phi,m) |
| $\Pi(\phi,\nu\mid m)\equiv\int_0^\phi \frac{d\theta}{(1-\nu\sin^2\theta)\sqrt{1-m\sin^2\theta}}$ | mpEllipticPi(PHI,NU,M) | mpelPi(phi,nu,m) |
| Jacobi's second form | | |
| $\mathcal{E}(u\mid m)\equiv\int_0^u \mathrm{dn}^2(t\mid m)\,dt$ | mJacobiEpsilon(U,M) | mjepsilon(u,m) |
| $\Lambda(u,\nu\mid m)\equiv\int_0^u \frac{dt}{1-\nu\,\mathrm{sn}^2(t\mid m)}$ | mJacobiLambda(U,NU,M) | mjlambda(u,nu,m) |

**Table A4.** Complete elliptic integrals. $-\infty < m \leq 1$

| Definition | Matrix argument | Scalar argument |
|---|---|---|
| $B(m) \equiv \mathbf{B}(1\|m) = B(\pi/2\|m)$ | mEllipticB(M) | melB(m) |
| $C(m) \equiv \int_0^1 \frac{t^2(1-t^2)\,dt}{\sqrt{(1-t^2)(1-mt^2)^3}} = \int_0^{\pi/2} \frac{\sin^2\theta\cos^2\theta\,d\theta}{\sqrt{(1-m\sin^2\theta)^3}}$ | mEllipticC(M) | melC(m) |
| $D(m) \equiv \mathbf{D}(1\|m) = D(\pi/2\|m)$ | mEllipticD(M) | melD(m) |
| $E(m) \equiv \mathbf{E}(1\|m) = E(\pi/2\|m) = \mathcal{E}(K(m)\|m)$ | mEllipticE(M) | melE(m) |
| $K(m) \equiv \mathbf{F}(1\|m) = F(\pi/2\|m)$ | mEllipticK(M) | melK(m) |
| $\Pi(\nu\|m) \equiv \mathbf{\Pi}(1,\nu\|m) = \Pi(\pi/2,\nu\|m) = \Lambda(K(m),\nu\|m)$ | mEllipticPi(NU,M) | melPi(nu,m) |

**Table A5.** Complementary complete elliptic integrals $0 \leq m < \infty$

| Definition | Matrix arguments | Scalar arguments |
|---|---|---|
| $E'(m) \equiv E(1-m)$ | mEllipticCE(M) | melCE(m) |
| $K'(m) \equiv K(1-m)$ | mEllipticCK(M) | melCK(m) |
| $\Pi'(\nu\|m) \equiv \Pi(\nu\|1-m)$ | mEllipticCPi(NU,M) | melCPi(nu,m) |

**Table A6.** Related functions. $-\infty < m \leq 1$

| Definition | Matrix arguments | Scalar arguments |
|---|---|---|
| $\mathbf{Z}(u\|m) \equiv \mathcal{E}(u\|m) - \frac{E(m)}{K(m)}u$ | mJacobiZeta(U,M) | mjzeta(u,m) |
| $Z(\phi\|m) \equiv E(\phi\|m) - \frac{E(m)}{K(m)}F(\phi\|m)$ | mpJacobiZeta(PHI,M) | mpjzeta(phi,m) |
| $\mathbf{\Omega}(u,\nu\|m) \equiv \mathbf{\Pi}(u,\nu\|m) - \frac{\Pi(\nu\|m)}{K(m)}u$ | mJacobiOmega(U,NU,M) | mjomega(u,nu,m) |
| $\Omega(\phi,\nu\|m) \equiv \Pi(\phi,\nu\|m) - \frac{\Pi(\nu\|m)}{K(m)}F(\phi\|m)$ | mpJacobiOmega(PHI,NU,M) | mpjomega(phi,nu,m) |
| $\Lambda_0(\beta\|m) \equiv \frac{m'\sin(2\beta)\Pi\left(\frac{m}{1-m'\sin^2\beta}\|m\right)}{\pi\sqrt{1-m'\sin^2\beta}}$ | mHeumanLambda(BETA,M) | mhlambda(beta,m) |

**Table A7.** Jacobian elliptic functions. $-\infty < m < \infty$

| Definition | Matrix arguments | Scalar arguments |
|---|---|---|
| $\mathrm{am}(x\|m) \equiv \int_0^x \mathrm{dn}(t\|m)dt$ | mJacobiAM(X,M) | mjam(x,m) |
| $\mathrm{cd}(x\|m) \equiv \mathrm{cn}(x\|m)/\mathrm{dn}(x\|m)$ | mJacobiCD(X,M) | mjcd(x,m) |
| $\mathrm{cn}(x\|m) \equiv \cos(\mathrm{am}(x\|m))$ | mJacobiCN(X,M) | mjcn(x,m) |
| $\mathrm{cs}(x\|m) \equiv \mathrm{cn}(x\|m)/\mathrm{ds}(x\|m)$ | mJacobiCS(X,M) | mjcs(x,m) |
| $\mathrm{dc}(x\|m) \equiv \mathrm{dn}(x\|m)/\mathrm{cn}(x\|m)$ | mJacobiDC(X,M) | mjdc(x,m) |
| $\mathrm{dn}(x\|m) \equiv \sqrt{1 - m\sin^2(\mathrm{am}(x\|m))}$ | mJacobiDN(X,M) | mjdn(x,m) |
| $\mathrm{ds}(x\|m) \equiv \mathrm{dn}(x\|m)/\mathrm{sn}(x\|m)$ | mJacobiDS(X,M) | mjds(x,m) |
| $\mathrm{nc}(x\|m) \equiv 1/\mathrm{cn}(x\|m)$ | mJacobiNC(X,M) | mjnc(x,m) |
| $\mathrm{nd}(x\|m) \equiv 1/\mathrm{dn}(x\|m)$ | mJacobiND(X,M) | mjnd(x,m) |
| $\mathrm{ns}(x\|m) \equiv 1/\mathrm{sn}(x\|m)$ | mJacobiNS(X,M) | mjns(x,m) |
| $\mathrm{sc}(x\|m) \equiv \mathrm{sn}(x\|m)/\mathrm{cn}(x\|m)$ | mJacobiSC(X,M) | mjsc(x,m) |
| $\mathrm{sd}(x\|m) \equiv \mathrm{sn}(x\|m)/\mathrm{dn}(x\|m)$ | mJacobiSD(X,M) | mjsd(x,m) |
| $\mathrm{sn}(x\|m) \equiv \sin(\mathrm{am}(x\|m))$ | mJacobiSN(X,M) | mjsn(x,m) |

**Table A8.** Inverse of Jacobian elliptic functions. $-\infty < m < \infty$

| Definition | Matrix arguments | Scalar arguments |
|---|---|---|
| $\mathrm{am}^{-1}(x \mid m) \equiv F(x \mid m)$ | mInverseJacobiAM(X,M) | mijam(x,m) |
| $\mathrm{cd}^{-1}(x \mid m) \equiv \mathbf{F}\left(\sqrt{\frac{1-x^2}{1-mx^2}} \mid m\right)$ | mInverseJacobiCD(X,M) | mijcd(x,m) |
| $\mathrm{cn}^{-1}(x \mid m) \equiv \mathbf{F}\left(\sqrt{1-x^2} \mid m\right)$ | mInverseJacobiCN(X,M) | mijcn(x,m) |
| $\mathrm{cs}^{-1}(x \mid m) \equiv \mathbf{F}\left(\frac{1}{\sqrt{1+x^2}} \mid m\right)$ | mInverseJacobiCS(X,M) | mijcs(x,m) |
| $\mathrm{dc}^{-1}(x \mid m) \equiv \mathbf{F}\left(\sqrt{\frac{x^2-1}{x^2-m}} \mid m\right)$ | mInverseJacobiDC(X,M) | mijdc(x,m) |
| $\mathrm{dn}^{-1}(x \mid m) \equiv \mathbf{F}\left(\sqrt{\frac{1-x^2}{m}} \mid m\right)$ | mInverseJacobiDN(X,M) | mijdn(x,m) |
| $\mathrm{ds}^{-1}(x \mid m) \equiv \mathbf{F}\left(\frac{1}{\sqrt{m+x^2}} \mid m\right)$ | mInverseJacobiDS(X,M) | mijds(x,m) |
| $\mathrm{nc}^{-1}(x \mid m) \equiv \mathbf{F}\left(\sqrt{1-\frac{1}{x^2}} \mid m\right)$ | mInverseJacobiNC(X,M) | mijnc(x,m) |
| $\mathrm{nd}^{-1}(x \mid m) \equiv \mathbf{F}\left(\sqrt{\frac{x^2-1}{mx^2}} \mid m\right)$ | mInverseJacobiND(X,M) | mijnd(x,m) |
| $\mathrm{ns}^{-1}(x \mid m) \equiv \mathbf{F}\left(\frac{1}{x} \mid m\right)$ | mInverseJacobiNS(X,M) | mijns(x,m) |
| $\mathrm{sc}^{-1}(x \mid m) \equiv \mathbf{F}\left(\frac{1}{\sqrt{1+x^2}} \mid m\right)$ | mInverseJacobiSC(X,M) | mijsc(x,m) |
| $\mathrm{sd}^{-1}(x \mid m) \equiv \mathbf{F}\left(\frac{x}{\sqrt{1+x^2}} \mid m\right)$ | mInverseJacobiSD(X,M) | mijsd(x,m) |
| $\mathrm{sn}^{-1}(x \mid m) \equiv \mathbf{F}(x \mid m)$ | mInverseJacobiSN(X,M) | mijsn(x,m) |

**Table A9.** Lemniscate functions

| Definition | Matrix argument | Scalar argument |
|---|---|---|
| $\mathrm{cl}(x) \equiv \frac{\sqrt{2}}{2}\mathrm{cn}\left(x\sqrt{2}, \frac{\sqrt{2}}{2}\right)$ | GaussCL(X) | gcl(x) |
| $\mathrm{sl}(x) \equiv \frac{\sqrt{2}}{2}\mathrm{sd}\left(x\sqrt{2}, \frac{\sqrt{2}}{2}\right)$ | GaussSL(X) | gsl(x) |
| $\mathrm{cl}^{-1}(x) \equiv \int_x^1 \frac{dt}{\sqrt{1-t^2}}$ | InverseGaussCL(X) | igcl(x) |
| $\mathrm{sl}^{-1}(x) \equiv \int_0^x \frac{dt}{\sqrt{1-t^2}}$ | InverseGaussSL(X) | Igsl(x) |

**Table A10.** Gudermannian functions

| Definition | Matrix argument | Scalar argument |
|---|---|---|
| $\mathrm{gd}(x) \equiv \int_0^x \frac{dt}{\cosh t} = \tan^{-1}(\sinh x)$ | GudermannGD | gd(x) |
| $\mathrm{gd}^{-1}(x) \equiv \int_0^x \frac{dt}{\cos t} = \sinh^{-1}(\tan x)$ | InverseGudermannGD(X) | igd(x) |

**Table A11.** Jacobi theta functions $0 \le q \le 1$

| Function | Matrix argument | Scalar argument |
|---|---|---|
| $\vartheta_1(x,q) \equiv 2q^{1/4}\sum_{n=1}^{\infty}(-1)^n q^{n(n+1)} \sin\left[(2n+1)x\right]$ | JacobiTheta1(X,Q) | jtheta1(x,q) |
| $\vartheta_2(x,q) \equiv 2q^{1/4}\sum_{n=1}^{\infty} q^{n(n+1)} \cos\left[(2n+1)x\right]$ | JacobiTheta2(X,Q) | jtheta2(x,q) |
| $\vartheta_3(x,q) \equiv 1+2\sum_{n=1}^{\infty} q^{n^2} \cos(2nx)$ | JacobiTheta3(X,Q) | jtheta3(x,q) |
| $\vartheta_4(x,q) \equiv 1+2\sum_{n=1}^{\infty}(-1)^n q^{n^2} \cos(2nx)$ | JacobiTheta4(X,Q) | jtheta4(x,q) |

**Table A12.** Neville theta functions $0 \le q \le 1$, $0 \le m \le 1$

| Function | Matrix arguments | Scalar arguments |
|---|---|---|
| $\theta_c(x,q) \equiv \theta_2\left(\frac{\pi x}{2K(k(q))}, q\right) \Big/ \theta_2(0,q)$ | NevilleThetaC(X,Q) | nthetac(x,q) |
| $\theta_d(x,q) \equiv \theta_3\left(\frac{\pi x}{2K(k(q))}, q\right) \Big/ \theta_3(0,q)$ | NevilleThetaD(X,Q) | nthetad(x,q) |
| $\theta_n(x,q) \equiv \theta_4\left(\frac{\pi x}{2K(k(q))}, q\right) \Big/ \theta_4(0,q)$ | NevilleThetaN(X,Q) | nthetan(x,q) |
| $\theta_s(x,q) \equiv \frac{2K(k)}{\pi}\theta_1\left(\frac{\pi x}{2K(k(q))}, q\right) \Big/ \theta_1'(0,q)$ | NevilleThetaS(X,Q) | nthetas(x,q) |
| | | |
| $\theta_c(x\mid m) \equiv \theta_c\left(\frac{\pi x}{2K(m)}, q(m)\right)$ | mNevilleThetaC(X,M) | mnthetac(x,m) |
| $\theta_d(x\mid m) \equiv \theta_d\left(\frac{\pi x}{2K(m)}, q(m)\right)$ | mNevilleThetaC(X,M) | mnthetac(x,m) |
| $\theta_n(x\mid m) \equiv \theta_n\left(\frac{\pi x}{2K(m)}, q(m)\right)$ | mNevilleThetaC(X,M) | mnthetac(x,m) |
| $\theta_s(x\mid m) \equiv \theta_s\left(\frac{\pi x}{2K(m)}, q(m)\right)$ | mNevilleThetaC(X,M) | mnthetac(x,m) |

**Table A13.** Related functions

| Function | Matrix arguments | Scalar arguments |
|---|---|---|
| $\text{nome}(k) \equiv \exp\left(-\frac{\pi K'(k)}{K(k)}\right)$ | EllipticNome(X,K) | elnome(k) |
| $\text{nome}(k) = q$ | InverseEllipticNome(Q) | ielnome(q) |

**Appendix B.** Program for calculation of the shape of the finite-strain cantilever under follower force

```
close all
% Data
psi1   = pi/3;  % angle between the direction of force an inward normal at free end
lambda = 10;    % generalized slenderness > 0
nu     = -1;    % stiffness ratio [-1..1]
omega  = 4;     % load factor
cantilever( psi1, lambda, nu, omega)

function cantilever( psi1, lambda, nu, omega)
% Cantilever under follower force ([1] Sec 5.1)
%
% Reference:
%  [1]  M.Batista - A closed-form solution for Reissner planar finite-strain
%       beam using Jacobi elliptic functions, International Journal of Solids
%       and Structures 87 (2016) 153–166

% Calculate constants
eta2   = (omega/lambda)^2;
k      = sin(psi1/2);                                   % Eq 53
m2     = nu*eta2*k^2/(1 - nu*eta2*(1 - k^2));           % Eq 34a
omega1 = omega*sqrt(1 + nu*eta2*(2*k^2 - 1));           % Eq 34b
k1     = sqrt((k^2 + m2)/(1 + m2));                     % Eq 39
C0     = -omega1 + elK(k1);                             % Eq 67
alpha  = 2*asin(k*jsn(C0,k1)/sqrt(1 + m2*jcn(C0,k1)^2));  % Eq 65

% Calculate shape
[ x, y, phi ] = shape( 0:0.01:1 );

% Calculate length of deformed beam
if nu == 1  % shearless beam only  Eq 46
    L = 1 - (1 + 2*k^2/m2)*eta2 + 2*k^2/m2/omega1*eta2*...
        (jlambda(omega1 + C0,m2/(1 + m2),k1) - jlambda(C0,m2/(1 + m2),k1));
else
    L = NaN;
end
fprintf('%s = %.16g\n','C',      C0)
fprintf('%s = %.16g\n','alpha', alpha)
fprintf('%s = %.16g\n','L',      L)
fprintf('%s = %.16g\n','X', x(end))
fprintf('%s = %.16g\n','Y', y(end))

% Plot shape
figure
hold on
% center line
plot(x,y,'k')
s = -1:0.1:1;
h = 1/lambda/2; % aux. beam thickness
% plot outer lines
plot(x + h*sin(phi),y - h*cos(phi),'k','LineWidth',2)
plot(x - h*sin(phi),y + h*cos(phi),'k','LineWidth',2)
% plot section
for n = 1:9
    [x0,y0,phi0] = shape(0.1*n);
    plot(x0 + h*sin(phi0)*s,y0 - h*cos(phi0)*s,'k');
end
% plot end sections
[x0,y0,phi0] = shape(0);
plot(x0 + h*sin(phi0)*s,y0 - h*cos(phi0)*s,'k','LineWidth',4);
[x0,y0,phi0] = shape(1);
plot(x0 + h*sin(phi0)*s,y0 - h*cos(phi0)*s,'k');
```

```
% plot force
s = -2:0.1:2;
quiver(x0 + h*sin(phi0),y0 - h*cos(phi0), -h*sin(phi0), h*cos(phi0),...
    4,'r','Linewidth',1.5);
txt = sprintf('%s = %g  %s = %g\n%s = %g  %s = %g  %s = %g%s',...
    '\lambda',lambda,'\nu',nu,'\omega',omega,'\kappa',kappa1,...
    '\psi',psi1*180/pi,'^{0}')
text(0.2,0.2,txt,'FontSize',14);
axis equal
grid on
hold off

    function [ x, y, phi ] = shape( s )
        C = C0*ones(size(s));
        xx = ((nu - 1)*omega^2*s/lambda^2)/ 2 + 2*omega1/(omega^2)*...
            ((EllipticE(k1) / EllipticK(k1) - 1/2)*omega1*s + ...
            JacobiZeta(omega1*s + C, k1) - JacobiZeta(C, k1) - ...
            (m2*(JacobiSN(omega1*s + C, k1).*JacobiCN(omega1*s + C, k1).*...
            JacobiDN(omega1*s + C, k1)./(1 + m2*JacobiCN(omega1*s + C, k1).^ 2)...
            - JacobiSN(C, k1).* JacobiCN(C, k1).*JacobiDN(C, k1)./...
            (1 + m2*JacobiCN(C, k1).^2))));     % Eq 43a
        yy = -(2*omega1*k*sqrt(1 + m2)/omega^2)*(JacobiCN(omega1*s + C,k1)./...
            (1 + m2*JacobiCN(omega1*s + C, k1).^ 2) -  JacobiCN(C, k1)./...
            (1 + m2*JacobiCN(C, k1).^ 2));      % Eq 43b
        x  =  xx*cos(alpha) + yy*sin(alpha);    % Eq 19a
        y  = -xx*sin(alpha) + yy*cos(alpha);    % Eq 19b
        phi = 2*asin(k*JacobiSN(omega1*s + C,k1)./...
            sqrt(1 + m2*JacobiCN(omega1*s + C,k1).^2)) - alpha;  % Eq 41, 22
    end

end
```